\documentclass[namedreferences]{solarphysics}
\usepackage[optionalrh]{spr-sola-addons} 
\usepackage{graphicx}        
\usepackage{amssymb}        
\usepackage{color}           
\usepackage{url}             
\usepackage[pdfborder={0 0 0 },urlcolor=blue,breaklinks]{hyperref}

\ifx \arxivurl  \undefined \def \arxivurl#1{\href{http://arxiv.org/abs/#1}{\textsf{#1}}}\fi 
\ifx \doiurl    \undefined \def \doiurl#1{\href{http://dx.doi.org/#1}{\textsf{#1}}}\fi 
\ifx \adsurl    \undefined \def \adsurl#1{\href{http://adsabs.harvard.edu/abs/#1}{\textsf{#1}}}\fi 

\newcommand{\aap}{    {\it Astron. Astrophys.}}

\newcommand{\apj}{    {\it Astrophys. J.}}

\newcommand{\mnras}{  {\it Mon. Not. Roy. Astron. Soc.}}

\newcommand{\solphys}{{\it Solar Phys.}}

\begin{document}

\begin{article}

\begin{opening}

\title{Sunspot Group Development in High Resolution}

\author{J.~\surname{Murak\"ozy}\sep
        T.~\surname{Baranyi}\sep
        A.~\surname{Ludm\'any}
       }

\runningauthor{J. Murak\"ozy \emph{et al.}}
\runningtitle{Sunspot Group Development}

   \institute{Heliophysical Observatory, Research Centre for Astronomy and Earth Sciences, Hungarian Academy of Sciences,
\\4010 Debrecen, P.O. Box 30, Hungary \\email: \url{murakozy.judit@csfk.mta.hu} 
email: \url{baranyi.tunde@csfk.mta.hu} email: \url{ludmany.andras@csfk.mta.hu}}

\begin{abstract}

The \textit{Solar and Heliospheric Obseratory/Michelson Doppler Imager--Debrecen Data} (SDD) sunspot catalogue provides an opportunity to study the details and development of sunspot groups on a large statistical sample. The SDD data allow, in particular, the differential study of the leading and following parts with a temporal resolution of 1.5 hours. In this study, we analyse the equilibrium distance of sunspot groups as well as the evolution of this distance over the lifetime of the groups and the shifts in longitude associated with these groups. We also study the asymmetry between the compactness of the leading and following parts, as well as the time-profiles for the development of the area of sunspot groups. A logarithmic relationship has been found between the total area and the distance of leading--following parts of active regions (ARs) at the time of their maximum area. In the developing phase the leading part moves forward; this is more noticeable in larger ARs. The leading part has a higher growth rate 
than the 
trailing part in most cases in the developing phase. The growth rates of the sunspot groups depend linearly on their maximum total umbral area. There is an asymmetry in compactness: the number of spots tends to be smaller, while their mean area is larger in the leading part at the maximum phase.

\end{abstract}
\keywords{sunspots, solar activity}
\end{opening}

\section{Introduction}
           \label{S-Introduction}

According to the generally accepted conception sunspot groups, or in more general terms solar active regions, emerge from large global toroidal magnetic fields generated at the bottom of the convective zone. The ideas about the causes of emergence are diverse, but it is also widely accepted that the process of emergence is driven by buoyancy and influenced by the strength, twist, and curvature of the flux tubes and the ambient velocity fields. The rising magnetic-flux ropes are mostly imagined to be ${\Omega}$-shaped formations; their top arches protrude from the convective zone and the intersections of the flux ropes with the photosphere are observed as sunspots. Mostly the directly observable surface properties of sunspot groups provide pieces of information about this complex process, \textit{e.g.} positions, sizes, developments, time profiles of rising and decaying phases, tilt angles, fragmentations, leading--following asymmetries, morphology, internal motions, cycle dependencies, and relations to the 
velocity fields.

The typical time profile of the sunspot group development has been known for a long time; its growing phase is usually shorter than its decaying phase. The two phases are governed by two different mechanisms. They were first examined both empirically and theoretically by \inlinecite{Cowling46}. He made calculations based on simple electromagnetic assumptions and obtained an expected decay time of about 300 years. The expected rise time was comparable to this result. These results indicated that these processes cannot be described by simple assumptions based on the conductivity of plasma. Further studies assumed different kinds of motion fields; a detailed summary of these mechanisms is given by \inlinecite{Fan09}.

Considerable efforts have been devoted to finding the most realistic theoretical description of the interaction between the magnetic and velocity fields resulting in those phenomena and processes which are directly observable at the surface. \inlinecite{Fan93} found possible theoretical reasons for the empirical finding that the subgroup of leading polarity tends to be more compact than the trailing part: they assume that the Coriolis force drives the flow in the rising flux from the leading part to the following one, and this was confirmed by \inlinecite{Abbett2001}.  \inlinecite{Fan94} also confirms that the magnetic field of the leading leg in the emerging loop is stronger than in the trailing leg. \inlinecite{Moreno94} and \inlinecite{Caligari95} found that the unstable flux tube ascends with a geometrical asymmetry: the leading leg is more inclined to the vertical direction than the trailing leg. \inlinecite{Caligari98} compared the consequences of two different initial conditions of buoyant ascent, the 
mechanical equilibrium \textit{vs.} temperature balance, and found that the resulting leading--following asymmetry is different in the two cases. Later three-dimensional work (\opencite{Abbett2000}, \citeyear{Abbett2001}) \href{Abbett2000} found further stabilising effects on the rising tubes from the initial magnetic field, its twist and curvature, as well as the rotation and convection. 

The theoretical works provide several features that may be observable at the surface, as is summarized by \inlinecite{Fisher2000a}. The present investigation focuses primarily on the developing phases of the sunspot groups until their largest extension.

\section{Statistical Study of Sunspot Group Details}
	\label{S-statistics}

      \subsection{The Observational Material} 
             \label{S-data}

The data of Solar and Heliospheric Obseratory/Michelson Doppler Imager--Debrecen sunspot Data (SDD) catalogue were used \cite{Gyori11}. This sunspot catalogue is more detailed than the Greenwich Photoheliographic Results (GPR) and its continuation, the Debrecen Photoheliographic Data (DPD). These traditional catalogues provide sunspot data on a daily basis and do not contain magnetic data. They are indispensable for long-term studies of the solar activity but the investigation of internal details in sunspot groups needs higher temporal resolution and also magnetic data. The SDD meets these requirements. The data of position, area and magnetic field for all observable sunspots and sunspot groups are given in 1\,--\,1.5 hour intervals. The catalogue covers the entire time interval of the SOHO/MDI mission: 1996\,--\,2010. 

In the present work, unless otherwise stated, the selection criteria of sunspot groups were as follows: only the positions between central meridian distances (CMD) of $\pm 60^{\circ}$ are considered, the group had to be observable on the solar disc for at least six days within this CMD range, it should have reached its maximum in this longitudinal range by requiring that at least two days after maximum were observed and the total area on the day after maximum is $10\,\% $ less than the maximum area. These criteria resulted in a sample of 390 sunspot groups.

    \subsection{Distance of Leading -- Following Subgroups}
             \label{S-distance}

\begin{figure}    
   \centerline{\includegraphics[width=0.35\textwidth,clip=,angle=-90]{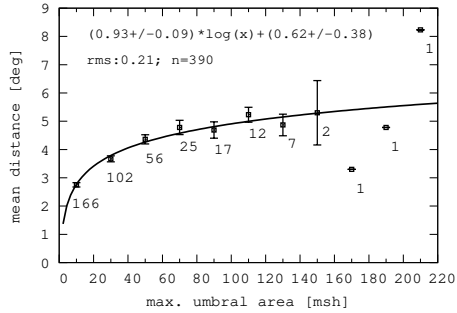}
                }
   
    \caption {Distances measured in heliographic degrees between leading and following polarity regions depending on the total umbral areas of sunspot groups measured in millionths of solar hemisphere (MSH). The means are plotted in the middle of the bins with the numbers of cases. The bins are 20 MSH wide. }

   \label{distances}
   \end{figure}

Following the emergence of the sunspots at the photosphere, the distance between the leading and following polarity parts grows in parallel with the growth of total spot area \cite{Gilman86}. The leading\,--\,following distance might be a parameter of the achieved state of maximum area at the time of the largest size.

Figure~\ref{distances} shows the relationship between the total umbral area of sunspot groups and the distance between their leading and following parts at the time of maximum area. The distance is computed between the ``center of mass" of both leading and following parts taking the umbral areas as masses. The mean distances with their errors have been computed in selected bins, their width was 20 millionths of solar hemisphere (MSH). The number of cases are indicated at each bin. Figure~\ref{distances} shows that there is a clear logarithmic relationship between the maximum area reached and the distance between leading and following spots; the function is indicated in Figure~\ref{distances}. The bigger the group in its most developed state, the stronger the stretching effect. This may indicate the role of magnetic tension in forming the longitudinal extension of the group.

  \subsection{Longitudinal Shifts}
             \label{S-shift}

\begin{figure}[H]    
   \centerline{\includegraphics[width=0.7\textwidth,clip=,angle=-90]{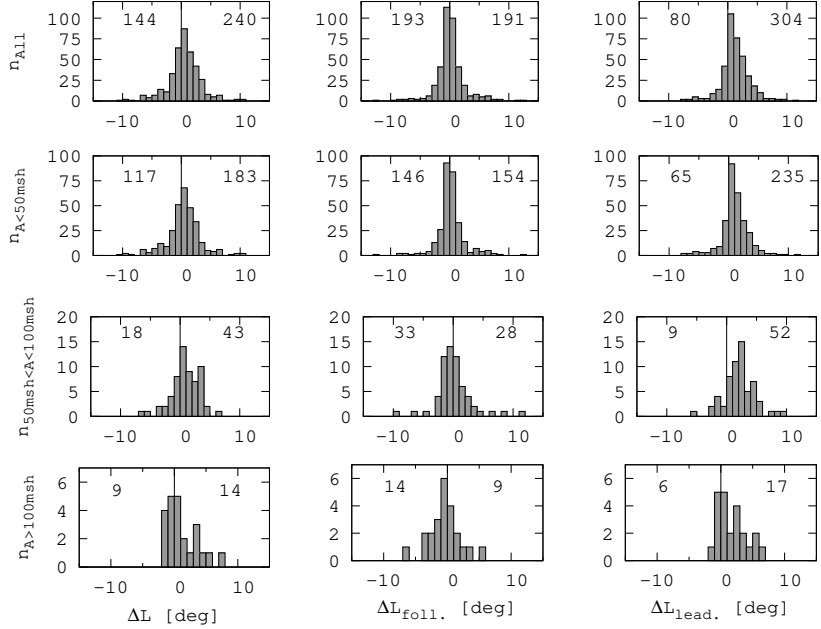}}
   
    \caption {Distributions of the longitudinal shifts between the first appearance and maximum  measured in heliographic degrees. The rows correspond to the sizes of the groups measured in MSH; the upper row shows the distributions of all cases, the other three rows show the distributions of three groups of sizes separately. The columns distinguish between the parts of active regions, left column: entire groups, middle and right columns: following and leading parts of the sunspot groups. The numbers in the negative and positive halves show the cases of forward and backward motions.}

   \label{shift}
   \end{figure}

After their emergence, the sunspot groups move in the longitudinal direction. The shifts from the first appearance until the maximum state have been computed and plotted in Figure~\ref{shift}. The left column of the figure shows the longitudinal shifts for the total groups and groups of different sizes, namely $A<$50, 50$<A<$100 and 100$<A$ MSH. The central and right columns show the same data for the following and leading parts of the groups. The numbers in the positive and negative domains indicate the cases of forward and backward shifts.

\begin{figure}[H]
   \centerline{\includegraphics[width=0.3\textwidth,clip=,angle=-90]{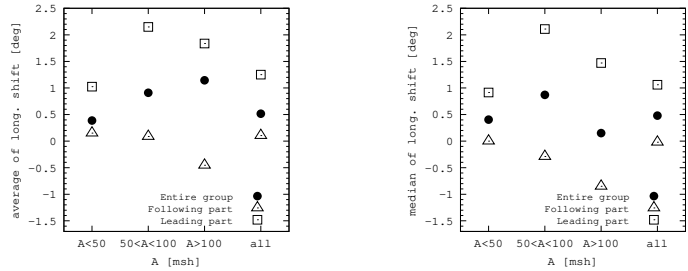}}
  
   \caption {Left panel: average of longitudinal shifts in the three ranges of sizes and the entire sample for the leading and following parts and the entire group. Right panel: the same as in the left panel for the medians of longitudinal shifts.}

   \label{1table}
   \end{figure}

The majority of the groups tend to move forward, but the backward and forward shifts depend on the size of the group. The upmost panels of Figure~\ref{shift} show all cases, here the forward motion is predominant in the leading parts ($80\,\%$ of all cases) and less significant in the following parts ($50\,\%$). Instead of the numbers of positive and negative cases, the means and medians of the shifts of the subgroups are more informative, see Table~\ref{T-shift} and in graphical representation Figure~\ref{1table}. In the smallest groups ($A<$50MSH) the forward motion of the leading parts until the maximum is small, its mean value is about one heliographic degree, the mean shift of the following parts is almost negligible. Thus the smallest groups remain very close to the position of first appearance; their mean shift is about 0.4 degrees. The following part in the middle group also does not move, and in the largest group slightly recedes. The forward motion of the entire group is most clearly expressed in the two 
largest groups; the 
means of these forward shifts are 0.909$\pm$0.302 (for sizes between 50$<A<$100 MSH) and 1.145$\pm$0.482 (for sizes of $A>$100 MSH). One can conclude that the diverging motion of the two parts is mainly produced by the forward motion of the leading part.

\begin{table}
\caption {Averages (upper table) and medians (lower table) of longitudinal shifts in heliographic degrees for sunspot groups of different umbral areas [$A$] indicated in MSH.}
\label{T-shift}
\begin{tabular}{lrrr}
  \hline
sample          & entire groups  & following parts &  leading parts \\ 
  \hline
all             &    0.516       &  0.104          &  1.251         \\
$A<$50          &    0.388       &  0.150          &  1.024         \\
50$<A<$100    &    0.909       &  0.087          &  2.149         \\
100$<A$         &    1.145       & -0.455          &  1.836         \\
  \hline
all             &    0.480      &  -0.020          &  1.060         \\
$A<$50          &    0.405      &   0.000          &  0.915         \\
50$<A<$100    &    0.870      &  -0.290          &  2.110         \\
100$<A$         &    0.150      &  -0.850          &  1.470         \\
  \hline
\end{tabular}
\end{table}

 \subsection{Growth Rate}
             \label{S-growth}

\begin{figure}    
   \centerline{\includegraphics[width=0.23\textwidth,clip=,angle=-90]{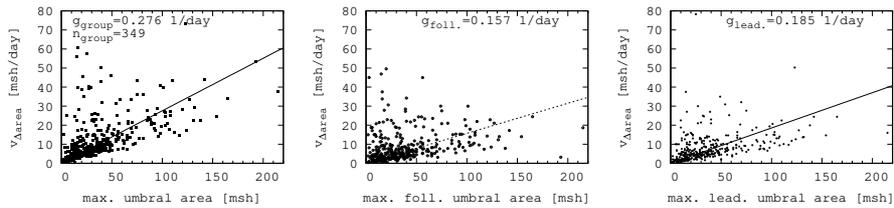}}
  
   \caption {Growth rates [\textit{g}] of sunspot groups depending on their maximum umbral area measured in MSH; {\it g} means the daily growth expressed as a fraction of the total umbral area. Left panel: dependence of the growth rates on the maximum umbral area of the entire groups, middle and right panels: the same relationships in the following and leading parts.}

   \label{growth1}
   \end{figure}

The growing phases of leading and following parts also show differences in time. The growth rates were determined in a fairly straightforward way, the maximum value of the total area was divided by the time interval between the  first appearance and maximum area of the group. The results are depicted in Figure~\ref{growth1}, for the entire group (left panel) and the following and leading parts (middle and right panels). The horizontal axes show the total umbral areas of the entire groups and the following and leading parts respectively, the rise times refer to the maxima of the relevant subgroups. It can be seen that larger groups grow faster and the leading growth rate is higher than the following one. The most important property is that the growth rate depends on the maximum area linearly. The present method is a simple procedure for the estimation of the growth rate; we will return to this relationship by using the temporal profiles derived in Section~\ref{S-profiles}.

\subsection{Asymmetric Compactness}
	\label{S-compact}

\begin{figure}    
   \centerline{\includegraphics[width=0.4\textwidth,clip=,angle=-90]{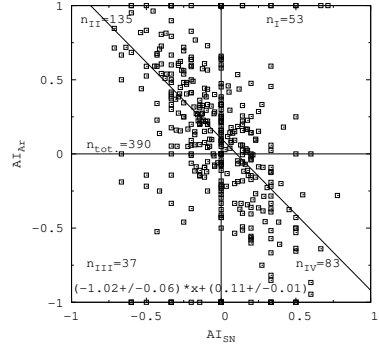}
               }
   
   \caption {Relation between leading/following asymmetry indexes of numbers ($AI_\mathrm{SN}$) and mean areas ($AI_\mathrm{Ar}$) of umbrae in sunspot groups.}

   \label{asymm1}
   \end{figure}

\begin{figure}    
   \centerline{\includegraphics[width=0.35\textwidth,clip=,angle=-90]{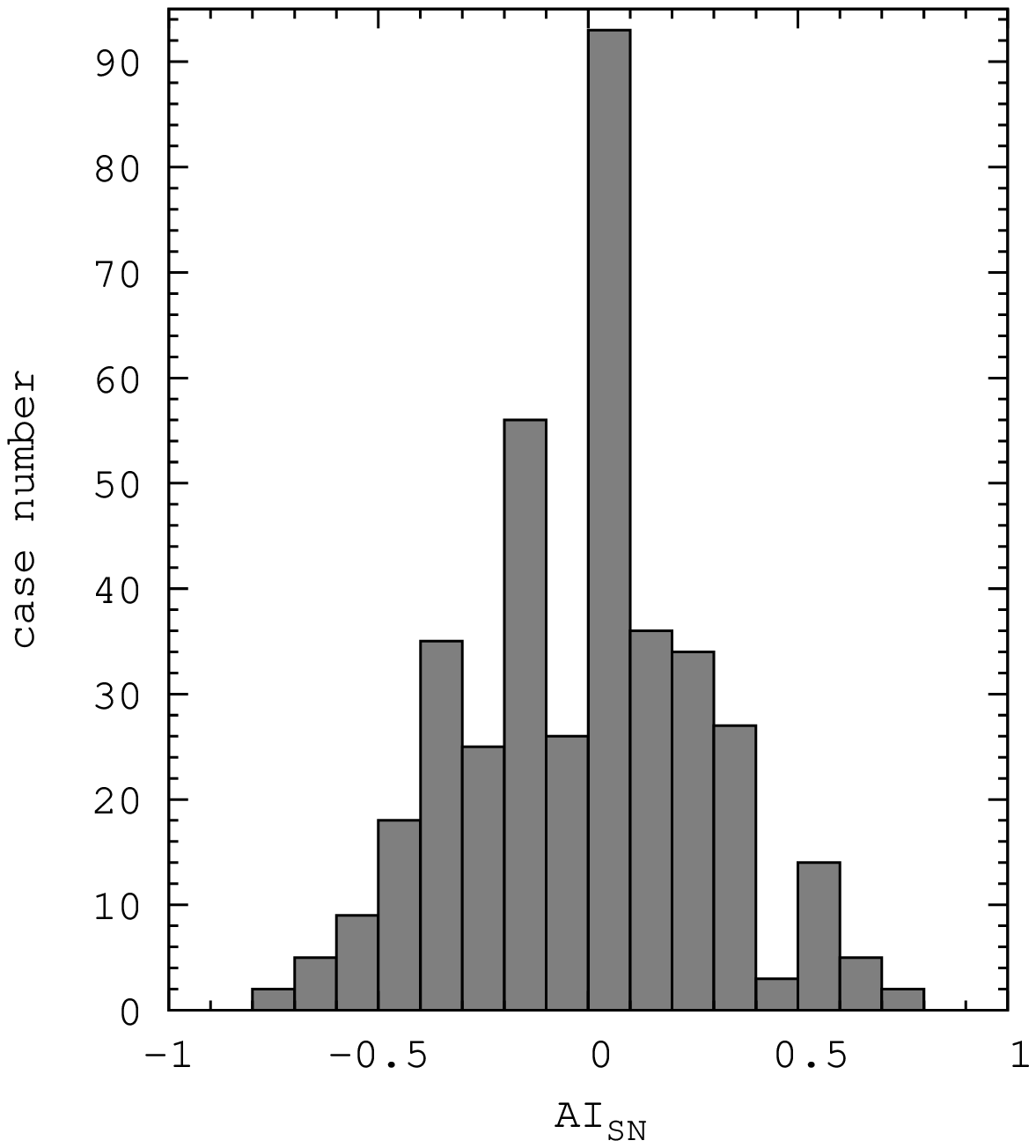}
               \hspace*{0.1\textwidth}
               \includegraphics[width=0.35\textwidth,clip=,angle=-90]{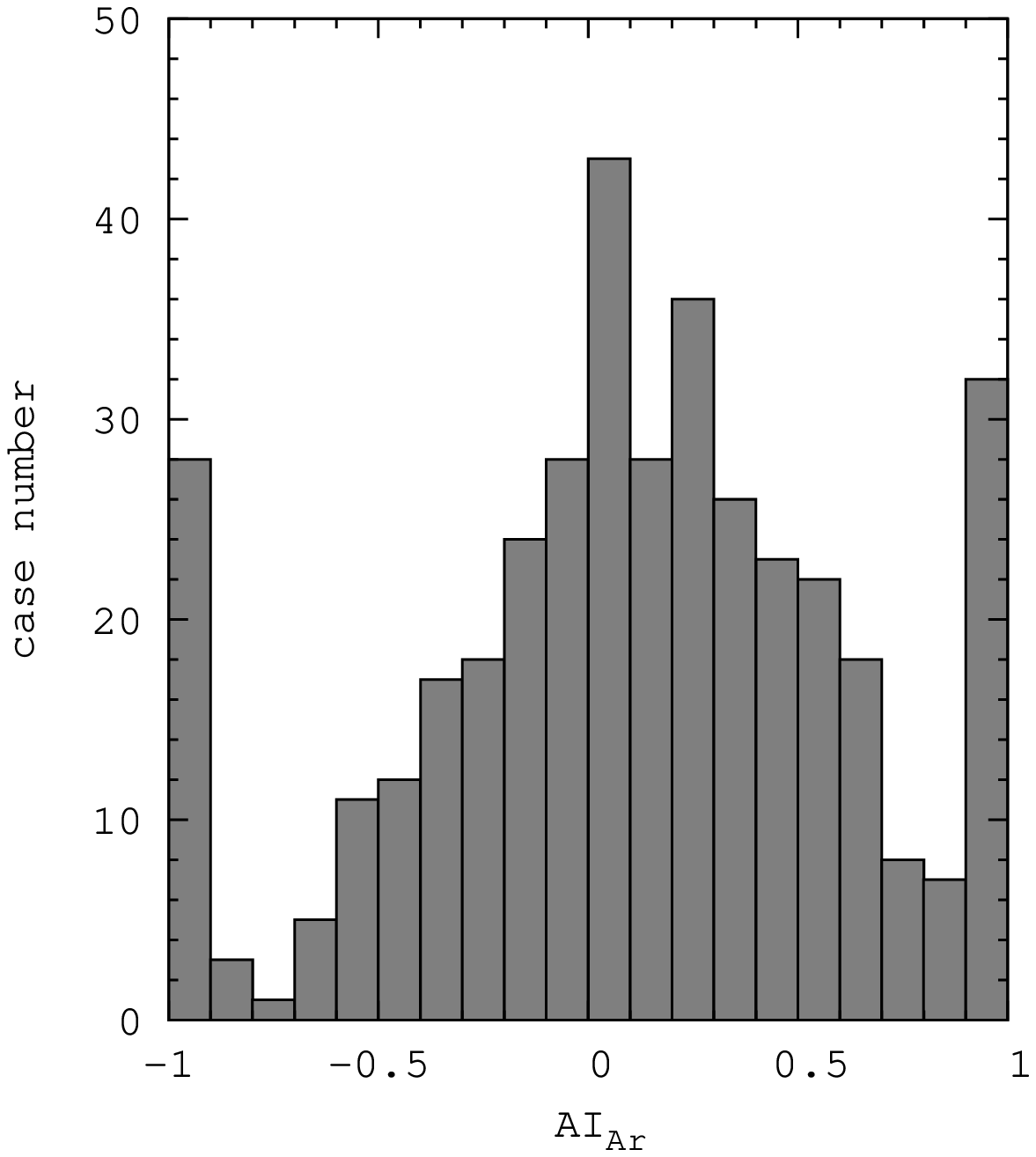}
               }
   
   \caption { Histograms of the leading--following asymmetry indexes. Left panel: numbers of umbrae ($AI_\mathrm{SN}$), right panel: mean areas of umbrae ($AI_\mathrm{Ar}$).}

   \label{asymm1a}
   \end{figure}

The levels of compactness in the leading and following parts of the sunspot groups are usually different. The asymmetry indexes [\textit{AI}] of sunspot groups were computed for the spot numbers [\textit{SN}] in the leading [${SN}_\mathrm{L}$] and following [${SN}_\mathrm{F}$] parts with the formula:

\begin{equation}
      AI_{\mathrm{SN}} = \frac{SN_{\mathrm{L}} - SN_\mathrm{F}}{SN_\mathrm{L} + SN_\mathrm{F}}
\label{asymfunc}
\end{equation}

\noindent
and the same formula was used for the computation of the asymmetry index [${AI}_\mathrm{Ar}$] of average sunspot areas [{\it Ar}] in the leading and following parts. Both parameters were considered at the time of the maximum of the sunspot-group development and only the umbrae were considered.  Figure~\ref{asymm1} shows the relation between the asymmetry indices $AI_\mathrm{SN}$ and $AI_\mathrm{Ar}$. Figure~\ref{asymm1a} shows the histograms of both asymmetry indexes for umbrae.

The trend of the diagrams shows that the majority of the cases belongs to the upper left quarter, \textit{i.e.} in the typical distribution the leading part contains fewer spots than the following part but the average area is larger in the leading part. The distribution of the points in the diagram shows a linear relationship, the fitted regression line is as follows:

\begin{equation}
      AI_\mathrm{Ar} = (-1.02 \pm 0.06)AI_\mathrm{SN} + (0.11 \pm 0.01)
\label{compact}
   \end{equation}

The linear relationship between the two asymmetry indexes has a simple meaning: if the leading or following part of the sunspot group contains more spots than the other part, then the mean area of these spots is smaller. What is more important, however, the regression line intersects the vertical axis at $AI_\mathrm{Ar}=0.11$ which means that even if the number of spots is the same in the two parts, the area asymmetry index is positive. This offset means that the mean sunspot area is typically larger in the leading part by $25\,\% $ than in the following part. This can be considered as a mean measure of the asymmetric clustering of the high density magnetic flux. 

The number of nonzero cases is indicated in all quarters, they show that beside the predominant upper left quarter the cases in the other quarters cannot be neglected either. The dots in the lower-right quarter contribute to the linear relationship. $29\,\% $ of all cases are in the domains of the other diagonal; however, these cases cannot be considered as distinct configuration types, as they are simply members of the scatter around the regression line in Figure~\ref{asymm1}. In other terms, the compactness asymmetry is better represented by the offset of the Equation~(\ref{compact}) than by the numbers of cases in the domains of Figure~\ref{asymm1}.

\subsection{ Time Profiles of Sunspot Group Development}
	\label{S-profiles}

Development and decay are important characteristics of active--region dynamics. They are governed by different physical processes. The emergence is driven by buoyancy, while the decay results from the impact of turbulent erosion \cite{Petrovay99}. These processes can also be mixed during the development of the active region. \inlinecite{Hathaway08} could only follow the development curve of the total area of a sunspot group with the one-day resolution allowed by the GPR. The SDD enables us to investigate the heading and trailing parts separately in 1.5-hour resolution.

A list has been compiled for those sunspot groups that were observable from their first appearance to their decaying phase. This criterion is more strict than those formulated in Section~\ref{S-data} and therefore this sample is more restricted than that analysed above; it contains 223 groups. The other difference is that in this study the umbra plus penumbra (U+P) areas are considered because their variation is more smooth than that of the umbrae. The following asymmetric gaussian function has been fitted to the time series of their total area data:

   \begin{eqnarray}
        f(t)=H\mathrm{exp} {\left(-\frac{(t-t_\mathrm{M})^2}{D(1+A(t-t_\mathrm{M}))}\right)} 	
        \label{asym}
   \end{eqnarray}

\noindent
where \textit{H} and $t_\mathrm{M}$ are the height and time of the maximum, and \textit{D} and \textit{A} determine the width and asymmetry respectively. This formula is a somewhat modified version of our previously applied function \cite{Murakozy} and another formula applied by \inlinecite{Du}. Its advantage over the two--component bell function (\textit{i.e.} two half-gaussians for the ascending and descending phases) is that the heights and times of the maxima as well as the ascending and descending slopes can be obtained by appropriate fitting to the data and these parameters can be compared directly for the leading and following parts. For two\,--\,component gaussians the maximum should be determined separately. 

\begin{figure}    
   \centerline{\includegraphics[width=0.5\textwidth,clip=,angle=-90]{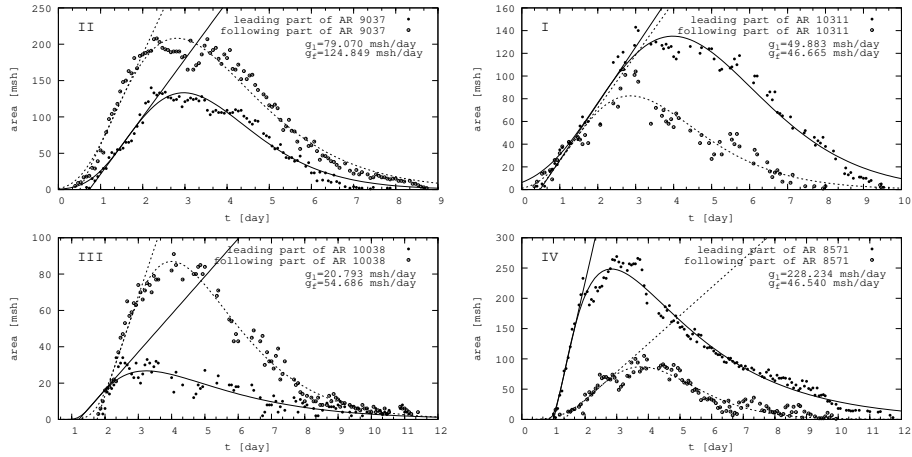}}
   
   \caption { Four cases for the relative time profiles of the leading and following parts in sunspot groups with the slopes of the fitted function in Equation~(\ref{asym}) at the inflection points.}

   \label{groups}
   \end{figure}

The time profiles of leading and following subgroups were treated separately. The following properties were examined: ratio of leading/following maxima [$H_{L}/H_{F}$],  difference between the times of leading and following maxima [$t_{L}-t_{F}$], the growth rates of leading and following parts, and the areal dependence of all of these data. In this case the growth rate was defined as the slope of the function in Equation~(\ref{asym}) at its inflection point. Concerning the relationships of leading--following maxima, the following cases were distinguished. The leading maximum can be: i) higher and later, ii) smaller and later, iii) smaller and earlier, iv) higher and earlier  than the following maximum. Figure~\ref{groups} shows one example for each case with the fitted Equation~(\ref{asym}) functions and the lines indicating the slopes at the inflection points.

\begin{figure}    
   \centerline{\includegraphics[width=0.5\textwidth,clip=,angle=-90]{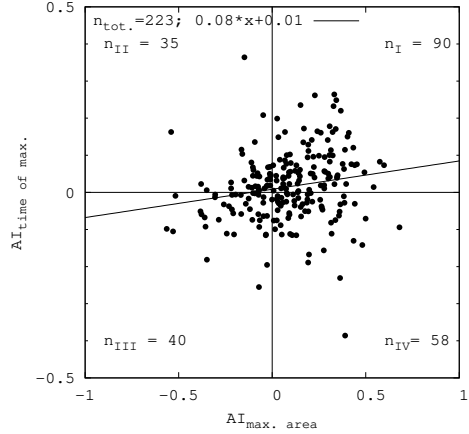}}
   
   \caption { Comparison of leading--following asymmetry indexes computed for the maximum areas and times of maxima. The number of cases is indicated in each segment.}

   \label{asymm2}
   \end{figure}

A further type of leading--following asymmetries can be studied by comparing the differences between the heights and times of maxima. Figure~\ref{asymm2} shows the comparison of the Equation~(\ref{asymfunc}) asymmetry indexes applied to the maximum areas and times of maxima. 

Figure~\ref{asymm2} shows that the most typical relative position of leading and following maxima is the later and higher maximum of the leading part (90 groups out of the sample of 223 groups) but none of the other three cases (35, 40, and 58 groups) can be neglected.

\begin{figure}    
   \centerline{\includegraphics[width=0.6\textwidth,clip=,angle=-90]{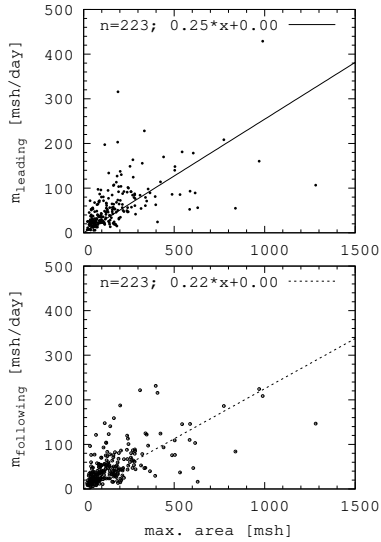}}
   
   \caption {Comparison of the growth rates in the leading and following parts of sunspot groups measured in $\mathrm{MSH\cdot day^{-1}}$. The data are obtained as the steepest slopes of the curves from Equation~(\ref{asym}) fitted to the temporal profiles of the  area data in the leading and following parts.}

   \label{growth2}
   \end{figure}

We also examine statistically the parameters of the curves separately. Figure~\ref{growth2} shows the areal dependence of the growth rates in the leading and following parts. In this case the growth rate is defined as the greatest steepness of the fitted Equation~(\ref{asym}) curve but the relationship between the fitted lines of Figure~\ref{growth2} is similar to that of Figure~\ref{growth1}, which is based on a comparatively simplified method and a larger sample. The leading and following growth rates are  0.185 \textit{vs.} 0.157 $\mathrm{day^{-1}}$ when using the method of Section~\ref{S-growth} whereas they are 0.25 \textit{vs.} 0.22 $\mathrm{day^{-1}}$ with the present time-profile analysis. This difference is of course due to the use of different methods; the recent method should give higher values because it considers the steepest moment of the development.

\begin{figure} 

  \centerline{\hspace*{0.015\textwidth   
              \includegraphics[width=0.35\textwidth,clip=,angle=-90]{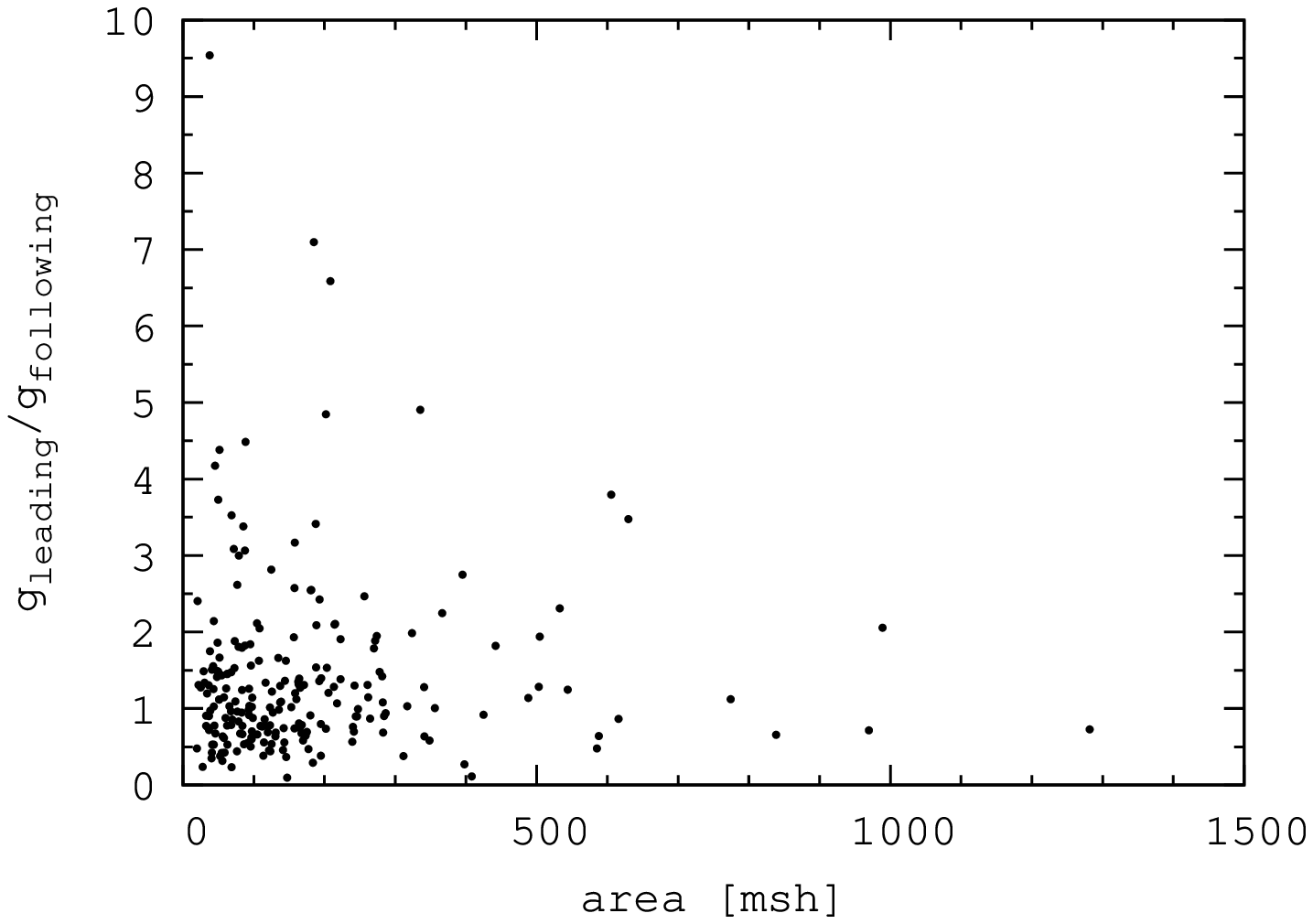}}
              \hspace*{-0.03\textwidth}
              \includegraphics[width=0.4\textwidth,clip=,angle=-180]{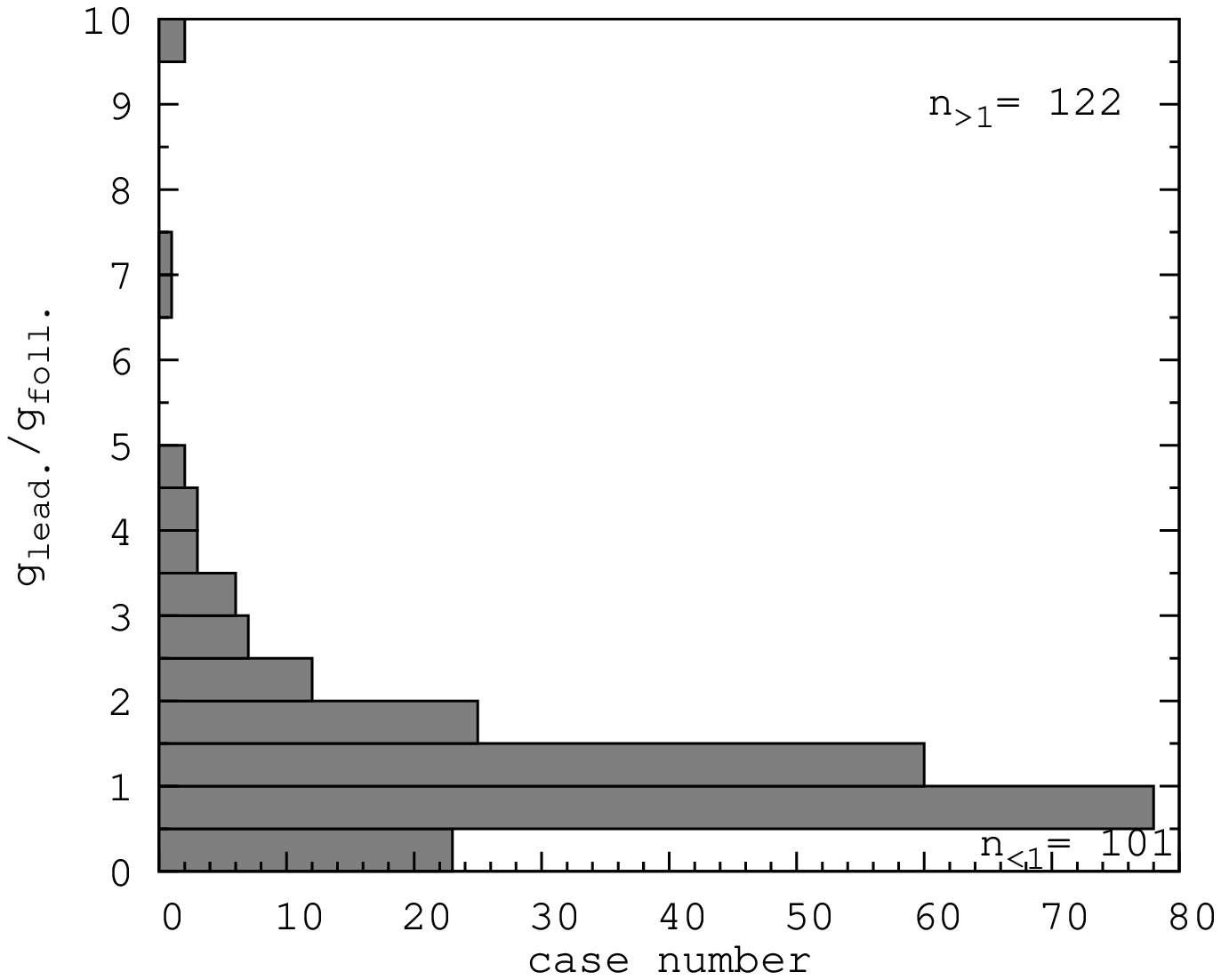}
              }  

  \centerline{\hspace*{0.015\textwidth   
              \includegraphics[width=0.35\textwidth,clip=,angle=-90]{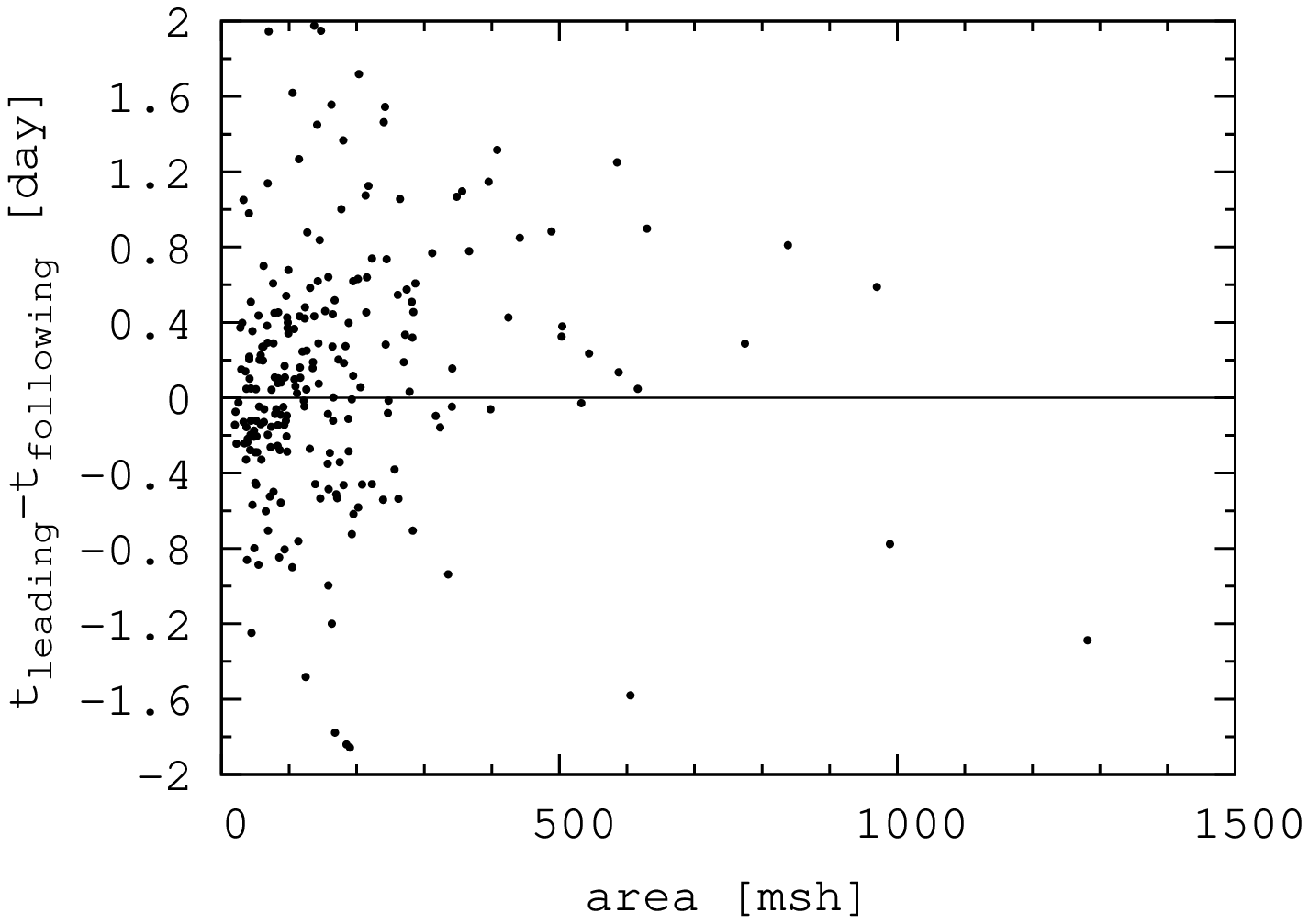}}
              \hspace*{-0.03\textwidth}
              \includegraphics[width=0.4\textwidth,clip=,angle=-180]{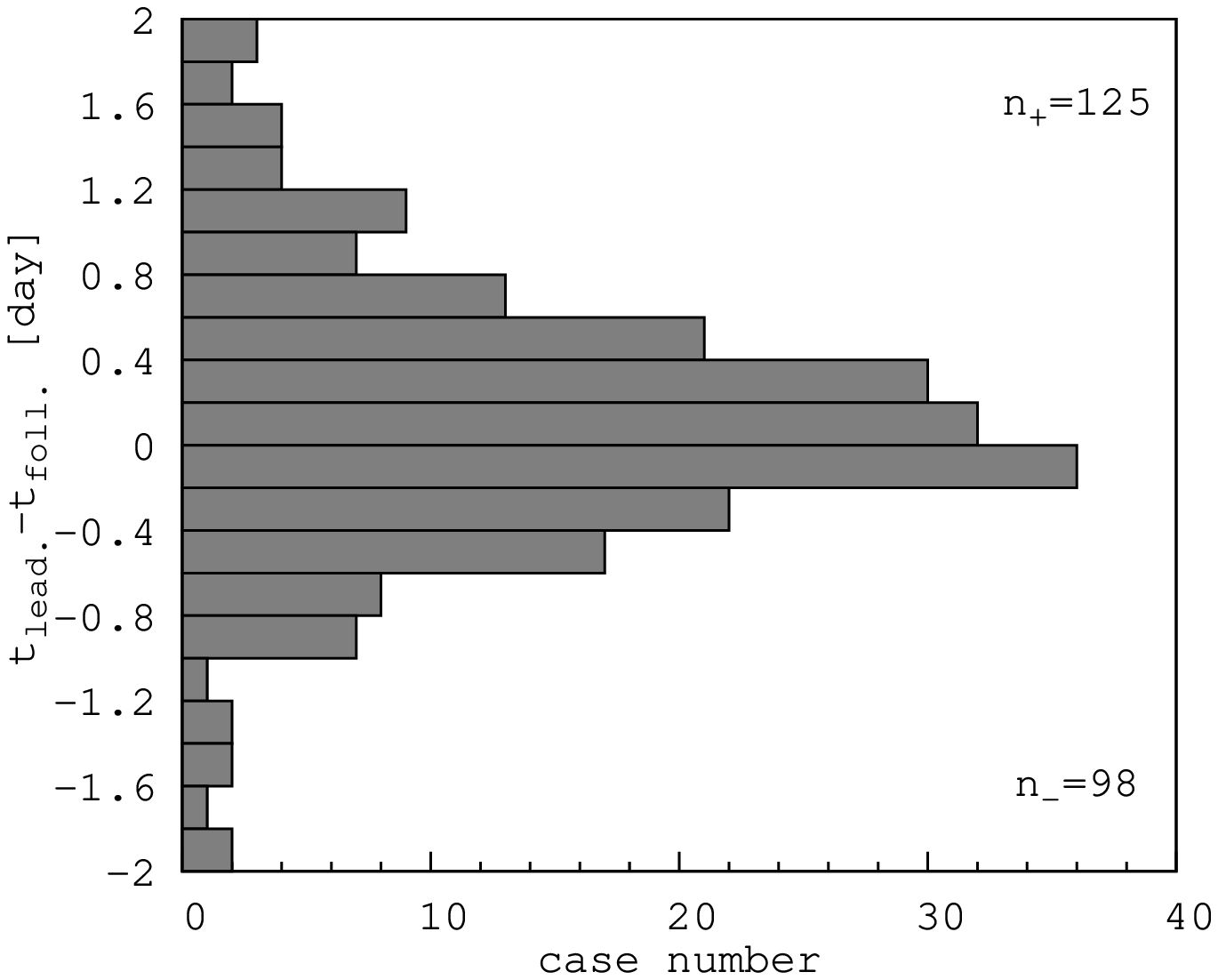}
              }  

  \centerline{\hspace*{0.015\textwidth   
              \includegraphics[width=0.35\textwidth,clip=,angle=-90]{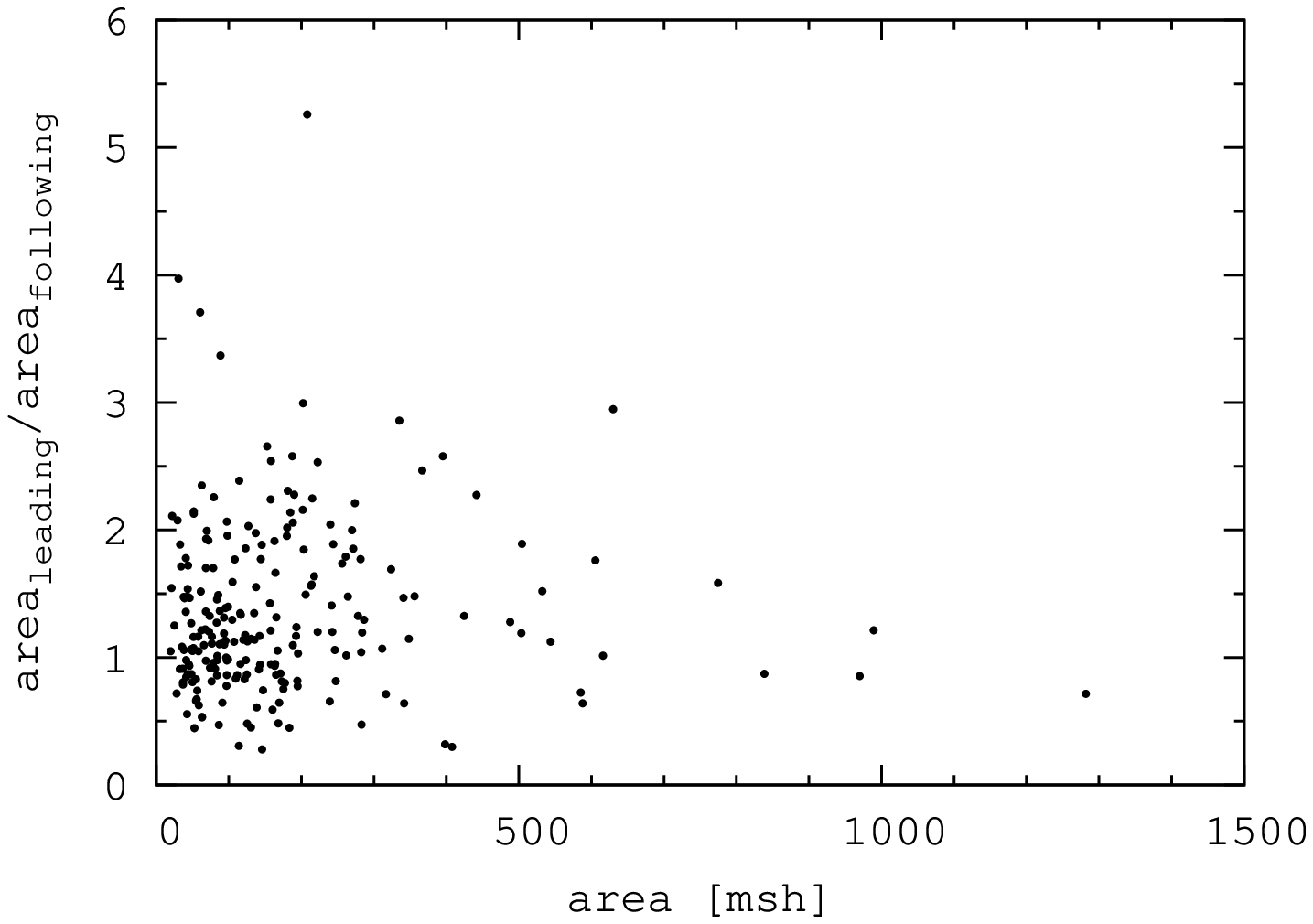}}
              \hspace*{-0.03\textwidth}
              \includegraphics[width=0.4\textwidth,clip=,angle=-180]{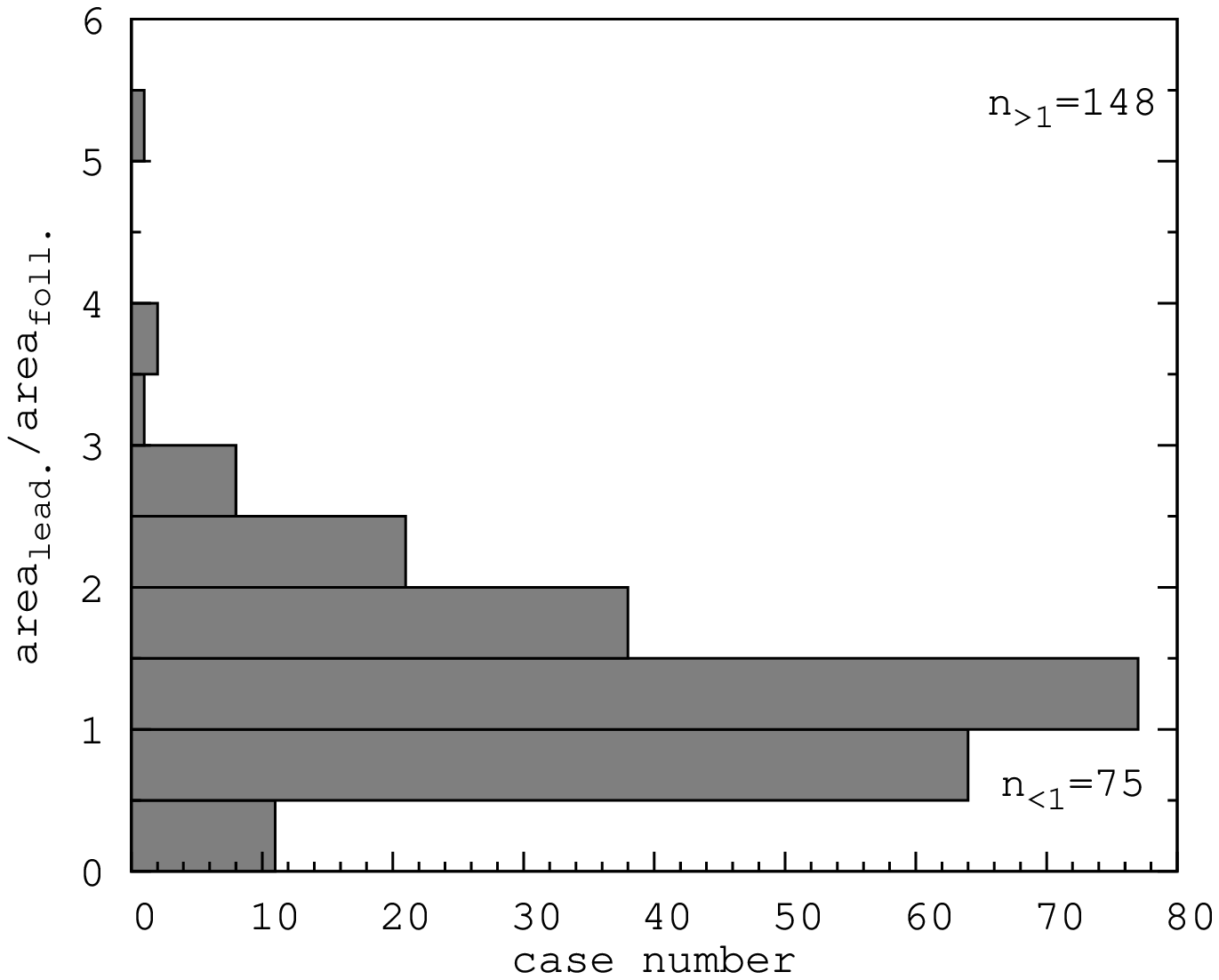}
              }
   
   \caption {Leading--following relationships of three parameters of the curves fitted using Equation~(\ref{asym}): the growth rates, the times and the values of maxima. Area dependencies (left panels) and histograms of cases (right panels) are shown for the following relationships: ratios of leading--following growth rates (upper row), differences of leading--following times of maxima (middle row), ratios of leading--following maxima (bottom row).}

   \label{area_dep}
   \end{figure}

In Figure~\ref{area_dep}, the area dependences and distributions are summarized for the relationships between the leading--following parameters of the fitted Equation~(\ref{asym}) curves. The area means the total U+P area of the group at maximum. None of the examined relationships depend on the area, but the right-hand panels show the histograms of the cases. The most unambiguous leading predominance is exhibited by the ratio of maximum areas; in two thirds of the cases the maximum area of the leading part is larger. The other two histograms are more intriguing, the leading/following ratio of growth rates is mostly larger than one, but the peak is just below one (upper panel). The difference between the times of maxima is mostly positive (the maximum of the leading part is later) but the peak of the histogram is just below zero.

\section{Discussion}
	\label{S-discuss}

Earlier investigations of sunspot--group developments used the classic photospheric data, the Mount Wilson observations \cite{Howard92a}, and their comparison with the Greenwich catalogue data \cite{Lustig94}. These works presented growth and decay rates of the sunspot groups, but with the restriction of the daily resolution and missing magnetic data, the results were presented on large statistical samples. The SDD catalogue offered several specific advantages for the present work to go into deeper detail of the processes. The first advantage is the temporal resolution  of 1.5 hours without nocturnal interruption, which allows us to follow the developments and motions with the precision that was necessary in the present work to determine the reliable temporal profiles of the developments and the times of maxima. The second advantage is the magnetic information of the spots, which makes the separation of leading and following parts much more reliable. Earlier works (\textit{e.g.} Howard, 1992b) had to separate 
the groups 
to spots at longitudes westward and eastward from the position of the area-weighted centroid of the group. This method may result in false separations in some cases. The other benefit of the polarity data is the reliable separation of entire sunspot groups. Considerable effort has been devoted in making the SDD to distinguish between two active regions emerging close to each other; this is not possible in the classic sunspot catalogues, and the unresolved cases distort the statistics. The third advantage is the availability of the data for both the sunspots and sunspot groups, which was indispensable for the present asymmetry studies.

Section~\ref{S-distance} presents results for the distances between the leading and following parts at the time of the maximum area of each sunspot group. According to Figure~\ref{distances} the growth of the mean leading--following distance is proportional to the logarithm of the maximum total umbral area of the group. This relationship may indicate an impact of the magnetic tension in stretching the sunspot group. The clarification of the role of the magnetic tension will need more detailed statistics of umbral areas along with the relevant flux--density data; this study is out of the scope of the present article and will be the topic of a subsequent work. Precedents of this study are sporadic in the literature and they treat the distance of the leading--following parts in the context of sunspot--group tilts without analytic formulation of its relationship to the maximum total size of the group. The plot presented by \inlinecite{Howard92b} (his Figure 8) weakly resembles our Figure~\ref{distances}, but in 
that study the separation of leading and following spots dispensed with the polarity data and the work was not restricted to the maximum state. \inlinecite{Tian99} presented a diagram about the relationship of the polarity separation and the total magnetic flux, regardless of the phase of development, but with emphasis on the tilts.

The results of Section~\ref{S-shift} show a leading--following asymmetry in the development. It is worth comparing the data of Figure~\ref{shift} and Table~\ref{T-shift} with the theoretical results of \inlinecite{Moreno94} and \inlinecite{Caligari95}, that the leading field lines are more inclined to the radial direction than the following ones. The measurements of \inlinecite{Gesztelyi90} and \inlinecite{Cauzzi98} about asymmetric magnetic fields with respect to the neutral line can also be explained by this property. The present data show that during the growth phase the following part mostly remains at the position of appearance (nearly vertically according to the results mentioned) whereas the leading part shifts forward and this is the source of the group lengthening. During this process, the inclination difference may decrease because \inlinecite{Cauzzi98} found decreasing asymmetry with the aging of the ARs. The larger is the group the larger is the forward shift of the leading part. The groups of smallest 
size (below 50 MSH) seem to remain at the starting position, \textit{i.e.} they proceed with the ambient plasma. This may be related to the theoretical finding of \inlinecite{Schussler05} about the dynamical disconnection of the emerged flux from the roots, this process may be more effective in the case of smaller groups.

Up to now this behaviour -- the growing polarity separation during the sunspot group development -- received little attention. Numerous theoretical works have been devoted to the magnetic-flux emergence -- their detailed overview is given by \inlinecite{Fan09} -- but in the photosphere they mostly focus on the tilt angles. The empirical works, however, were restricted by the lack of detailed datasets. The earliest works were case studies. Expansion velocities were reported by \inlinecite{Nagy} for one active region and by \inlinecite{Chou} for 24 bipolar regions but not restricted to the time interval prior to the maximum. \inlinecite{Howard89} presented polarity separation data for 7629 active regions regardless of their area and phase of development and without direct magnetic data. \inlinecite{Strous} presented mean diverging velocities of spots in a single active region but not restricted to the time interval between the birth and maximum. \inlinecite{Schussler97} investigated 3793 active regions over 108 
years and 
found that secondary groups tend to emerge westward from the earlier emerged primary ones in the case of large groups. Like our results in the Section~\ref{S-shift}, this can be interpreted by the result of \inlinecite{Moreno94} that a larger amount of emerging flux is more asymmetric, \textit{i.e.} it is more inclined to the radial direction in the western leg than in the eastern one. Nevertheless, the phenomenon of secondary groups is different from the shifts presented in Section~\ref{S-shift}, which apparently contains the first detailed data about the displacements of the opposite--polarity regions in developing sunspot groups.

Section~\ref{S-growth} presents growth rates of sunspot groups separately for the leading and following parts. Figure~\ref{growth1} shows the linear relationship between the maximum total area and the daily mean growth rate. For the entire group this is 27.6\,\% of the actual total area independently of the size. This linearity may give a contribution to the description of buoyant motion. The buoyant force is estimated to be $B_{0}^{2}/2\mu H_{p}$ \cite{Fan93} where $B_{0}$ is the initial value of the magnetic field and $H_{p}$ is the external pressure scale height. This means that stronger magnetic field can be expected to result in higher emerging velocity. Both the magnetic field and the emerging velocity are represented here by proxy measures, by the total umbral area, and the growth rate, but their linear proportionality implies that at the surface the emerging velocity is the same for all active-region magnetic fields independently of their sizes (the strengths of the magnetic fields) or their 
presumed emerging velocities within the convective zone. This can happen in such a way that even if the velocities are different during the emergence they might apparently be equalized before reaching the surface, perhaps by the different or varying drag force.

Section~\ref{S-compact} presents the asymmetry of compactness between the leading and following parts of sunspot groups. Figure~\ref{asymm1} shows that in the most typical cases the leading part contains less spots than the following part but their mean area is larger than that of the spots in the  following part. This is consistent with the results of \inlinecite{Fan93} that a significant asymmetry is produced in the rising magnetic flux rope by the Coriolis force; the leading part is more compact and the following part gets fragmented. Figure~\ref{asymm1} also shows that in the case of equal spot number in the two parts, the area asymmetry is $AI_{Ar}=0.11$  for umbrae. This means that on average 25\,\% larger area comprises high-density magnetic flux in the leading part than in the following part. The rest of the magnetic flux is dispersed in the ambient facular clusters. This may be compared to the result of \inlinecite{Yamamoto12}, who analysed the area asymmetries of the opposite polarity parts in active 
regions on the basis of magnetograms. His asymmetry parameter differs from our Equation~(\ref{asymfunc}):  $A=\mathrm{log}(S_\mathrm{F}/S_\mathrm{L})$, where $S_\mathrm{F}$ and $S_\mathrm{L}$ denote the areas of following and leading polarities. He found that the average area-asymmetry ratio was distributed between -0.2 and 0.4; the peak of the distribution was at 0.03. The conversion of this {$A$} = 0.03 value to the {\it AI} index defined by Equation~(\ref{asymfunc}) gives -0.037. By converting to percentages, one obtains that in the most typical cases the area of following polarity region is 7\,\% higher than that of the leading polarity. The direct comparison of this asymmetry data with ours cannot give reliable assessments about the amounts of magnetic fluxes within and out of the spots, because that work analysed the ARs at the centre of the solar disc whereas we considered them at their maximum state. 

Section~\ref{S-profiles} describes the statistics of the fitting parameters in the Equation~(\ref{asym}) fitted to the leading and following time profiles of the selected 223 sunspot groups, a sample of 446 curves. The comparison of Figures~\ref{growth1} and ~\ref{growth2} gives qualitatively similar results for the growth rates. The main point is the linear dependence on the total area as discussed above. Further statistical properties of the fitting parameters: the maximum area of the leading part is higher in two thirds of the cases, and the leading part reaches its maximum later than the following part in 56\,\% of the cases.

\section{Summary}
	\label{S-sum}

The obtained results can be summarized as follows.

i) The distance between the leading and following parts of the sunspot groups increases with increasing total area [{\it A}] measured in the most developed state of the group. This dependence is described by a logarithmic function and it may mean that magnetic tension plays a role in the longitudinal extent of the sunspot group.

ii) The dependence of the growth rate on the maximum umbral area is linear for the whole group as well as the leading and following parts; this was obtained by both the simplified method (Section~\ref{S-growth}) and the time-profile analysis (Section~\ref{S-profiles}). This linearity means that close to the surface the emerging speed is independent of the amount of emerging magnetic flux.

iii) The longitudinal shift of the whole group during the growth phase shows dependence on its total area but the following part mostly remains close to the starting location and the leading part shifts forward. The mean value of the shift of the leading part is about $\Delta L \approx 2^{\circ} $ until the maximum in the groups of maximum size exceeds 50 MSH.

iv) The following asymmetries have been found between the leading and following parts. In the state of maximum area the compactness is different: the leading part contains fewer and larger spots than the following part. The mean area of the leading spots is about 25\,\% larger than that of following spots. In two thirds of all cases, the maximum area of the leading part is higher than that of the following part. The development is also different in the two subgroups; in most cases the areal dependence of the growth rate is stronger in the leading part and the time of maximum is later in the leading part.

\begin{acks}

The research leading to these results has received funding from the European Space Agency project, ESA PECS 98081. The results have been presented at the NSO Workshop 26, Sunspot, New Mexico, and JM wishes to acknowledge the support provided by the organisers. The authors are grateful to the unknown referee for the substantial improvement of the manuscript.

\end{acks}

\end{article} 

\end{document}